\documentclass{iaus}
\usepackage{graphics}

  \checkfont{eurm10}
  \iffontfound
    \IfFileExists{upmath.sty}
      {\typeout{^^JFound AMS Euler Roman fonts on the system,
                   using the 'upmath' package.^^J}%
       \usepackage{upmath}}
      {\typeout{^^JFound AMS Euler Roman fonts on the system, but you
                   dont seem to have the}%
       \typeout{'upmath' package installed. iaus.cls can take advantage
                 of these fonts,^^Jif you use 'upmath' package.^^J}%
      }
  \else
  \fi

  \checkfont{msam10}
  \iffontfound
    \IfFileExists{amssymb.sty}
      {\typeout{^^JFound AMS Symbol fonts on the system, using the
                'amssymb' package.^^J}%
       \usepackage{amssymb}%
         
       \let\ge=\geqslant  
      }{}
  \fi

  \IfFileExists{amsbsy.sty}
    {\typeout{^^JFound the 'amsbsy' package on the system, using it.^^J}%
     \usepackage{amsbsy}}
    {\providecommand\boldsymbol[1]{\mbox{\boldmath $##1$}}}

\newcommand\kms{\nobreak\mbox{$\;$km\,s$^{-1}$}}
\newcommand\teff{\nobreak\mbox{T$_{\rm eff}$}}
\newcommand\mictrb{\nobreak\mbox{$\xi_{\rm t}$}}

\title[]{Observations of convection in A-type stars}

\author[B. Smalley]%
{Barry Smalley}

\affiliation{Astrophysics Group, School of Chemistry \& Physics, Keele
University, Staffordshire ST5 5BG, United Kingdom}

\pubyear{2004}
\volume{224}
\pagerange{1--8}
\date{?? and in revised form ??}
\jname{The A-Star Puzzle}
\editors{J. Zverko, W.W. Weiss, J. Ziznovsky \& S.J. Adelman, eds.}

\begin{document}
\maketitle

\begin{abstract} Convection and turbulence in stellar atmospheres have a
significant effect on the emergent flux from A-type stars. The recent
theoretical advancements in convection modelling have proved a challenge to the
observers to obtain measurements with sufficient precision and accuracy to
allow discrimination between the various predictions.

A discussion of the current observational techniques used to evaluate the
various convection theories is presented. These include filter photometry,
spectrophotometry, hydrogen lines, and metal lines. The results from these
techniques are given, along with the successes and limitations.

\end{abstract}

\firstsection 

\section{Introduction}

The gross properties of a star, such as broad-band colours and flux
distributions, are significantly affected by the effects of convection in stars
later than mid A-type. Consequently, our modelling of convection in stellar
atmosphere models can significantly alter our interpretation of observed
phenomena. 

Convection in stellar atmospheres is usually based on mixing-length theory
(MLT) of B\"{o}hm-Vitense (\cite{BV58}). In their discussion of the Kurucz
(\cite{KUR79}) {\sc atlas6} models Relyea \& Kurucz (\cite{RK78}) found
discrepancies between theoretical and observed $uvby$ colours which might be
the result of inappropriate treatment of convection within the models.
Subsequently, several attempts have been undertaken to improve the situation.
Lester et al. (\cite{LLK82}), for example, introduced ``horizontally
averaged opacity'' and a ``variable mixing length'' which improved the match
with observed $uvby$ colours, but did not remove all the discrepancies. The
{\sc atlas9} models introduced an ``approximate overshooting'' which has not
been without its critics (see Castelli et al. \cite{CGK97} for
details).

Canuto \& Mazzitelli (\cite{CM91,CM92}) proposed a turbulent model of
convection in order to overcome one of the most basic short-comings of MLT,
namely that a single convective element (or ``bubble'' or ``eddy'') responsible
for the transport of all the energy due to convection. This new model accounts
for eddies of various sizes that interact with each other. The CM convection
model was implemented in the {\sc atlas9} code by Kupka (\cite{KUP96}).

Theoretical studies of convection in A-type stars have suggested that
convection is inefficient (Freytag et al. \cite{FLS96}), thus has only a very
small influence on atmospheric structure (Heiter et al. \cite{HEI+02}).
However, these effects are nonetheless still important and in need of
observational confirmation. For the early A-type stars (hotter than
$\sim$8500~K) convective flux is negligible. It is only from mid A-type and
later when convection becomes important, as an extensive convection zone
develops in the photosphere and below (e.g. Weiss \& Kupka \cite{WK99}).

Convection in A-type stars poses a challenge to both theorists and observers.
In this review I will concentrate on the observational effects of convection
and what can be deduced from various types of observations.

\section{Colours}

Photometric indices are a fast and efficient method for determining approximate
atmospheric parameters of stars. For the commonly-used Str\"{o}mgren $uvby$
system a vast body of observational data exists (e.g. Hauck \& Mermilliod
\cite{HM98}) which can be used to estimate parameters using calibrated model
grids (e.g. Moon \& Dworetsky \cite{MD85}, Lester et al. \cite{LGK86},
Smalley \& Dworetsky \cite{SD95}). Conversely, knowing atmospheric parameters
from other methods, allows observed colours to be compared to model
predictions. This method has been used to compare various treatments of stellar
convection.

Smalley \& Kupka (\cite{SK97}) compared the predicted $uvby$ colours for the CM
model with that from the standard Kurucz (\cite{KUR93}) MLT models with and
without ``approximate overshooting''. Comparison against fundamental \teff\ and
$\log g$ stars revealed that the CM models gave better agreement than MLT
without overshooting. Models with overshooting were clearly discrepant. This
result was further supported by stars with \teff\ obtained from the
Infrared Flux Method (IRFM) and $\log g$ from stellar evolutionary models.
However, some discrepancies still remained, including a ``bump'' around 6500~K
in the $\log g$ obtained for the Hyades and continued problems with the
Str\"{o}mgen $m_0$ index. Similar conclusions were found by Schmidt
(\cite{SCH99}) using Geneva photometry.

\begin{figure}
 \includegraphics{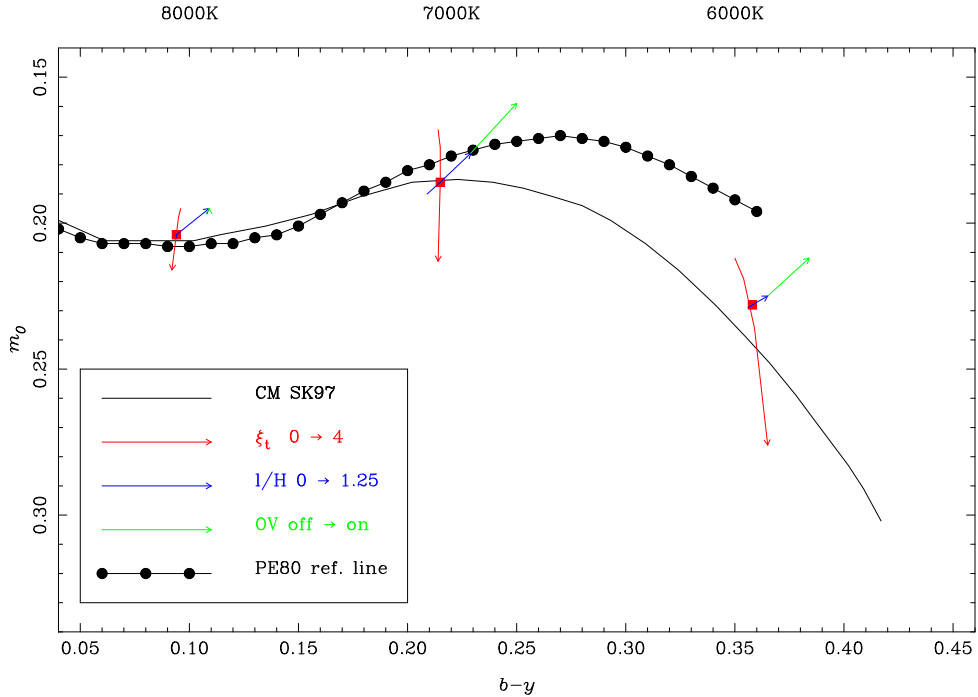}
 
 \caption{The variation of $m_0$ index with $b-y$ showing the sensitivity to
 microturbulence (\mictrb), mixing-length ($l/H$) and ``approximate
 overshooting''. At each temperature the model with $\log g$ = 4.0, \mictrb\ =
 2 \kms\ and $l/H$ = 0.5 is denoted by a square. The arrows indicate the effect
 of varying \mictrb, $l/H$ and including overshooting (for $l/H$ = 1.25). The
 Philip \& Egret (\cite{PE80}) main-sequence and the Smalley \& Kupka
 (\cite{SK97}) CM relationships are included for reference. Note that the
 squares do not lie on the CM relationship due to differing $\log g$ values.}

 \label{fig:m0}
\end{figure}

The $m_0$ index is sensitive to metallicity and microturbulence, but also
convection efficiency as discussed by Relyea \& Kurucz (\cite{RK78}) and
Smalley \& Kupka (\cite{SK97} Fig.~6). Inefficient convection (CM and MLT
$l/H$ $\sim$ 0.5) clearly works in the domain of the A stars down to $b-y$
$\sim$ 0.20 (approx 7000~K, F0). For cooler stars, convection becomes more
efficient and substantive within the atmosphere and higher values of
mixing-length and lower microturbulent velocities would be required to fit the
observed $m_0$ indices of the main sequence (see Fig.~\ref{fig:m0}). The 2d
numerical radiation hydrodynamics calculations of Ludwig et al. (\cite{LFS99})
are useful in this regard. They indicate a rise in mixing-length from $l/H$
$\sim$1.3 at 7000~K to $l/H$ = 1.6 for the Sun (5777~K), while a much lower
$l/H$ $\sim$ 0.5 was found for models at 8000~K (Freytag \cite{FRE95}). This is
in agreement with that implied by $m_0$ index. However, around 6000~K there
still remains a significant discrepancy, which could only be reduced by invoking
the approximate overshooting option (Fig.~\ref{fig:m0}). Thus, none of the
convection models used in classical model atmospheres allows for the
reproduction of the $m_0$ index, unless two parameters, i.e. $l/H$ and the
amount of ``approximate overshooting'', are varied over the H-R Diagram.

\section{Fluxes}

The observed stellar flux distribution is influenced by the effects of
convection on the atmospheric structure of the star. As we have seen with
photometric colours, these effects have a clearly observable signature. Hence,
high precision stellar flux measurements will provide significant and useful
information on convection.

Lester et al. (\cite{LLK82}) presented a study of convective model
stellar atmospheres using a modified mixing-length theory. They found small,
systematic differences in the optical fluxes. Their figures demonstrate that
convection can have a measurable effect on stellar fluxes.

\begin{figure}
 \includegraphics{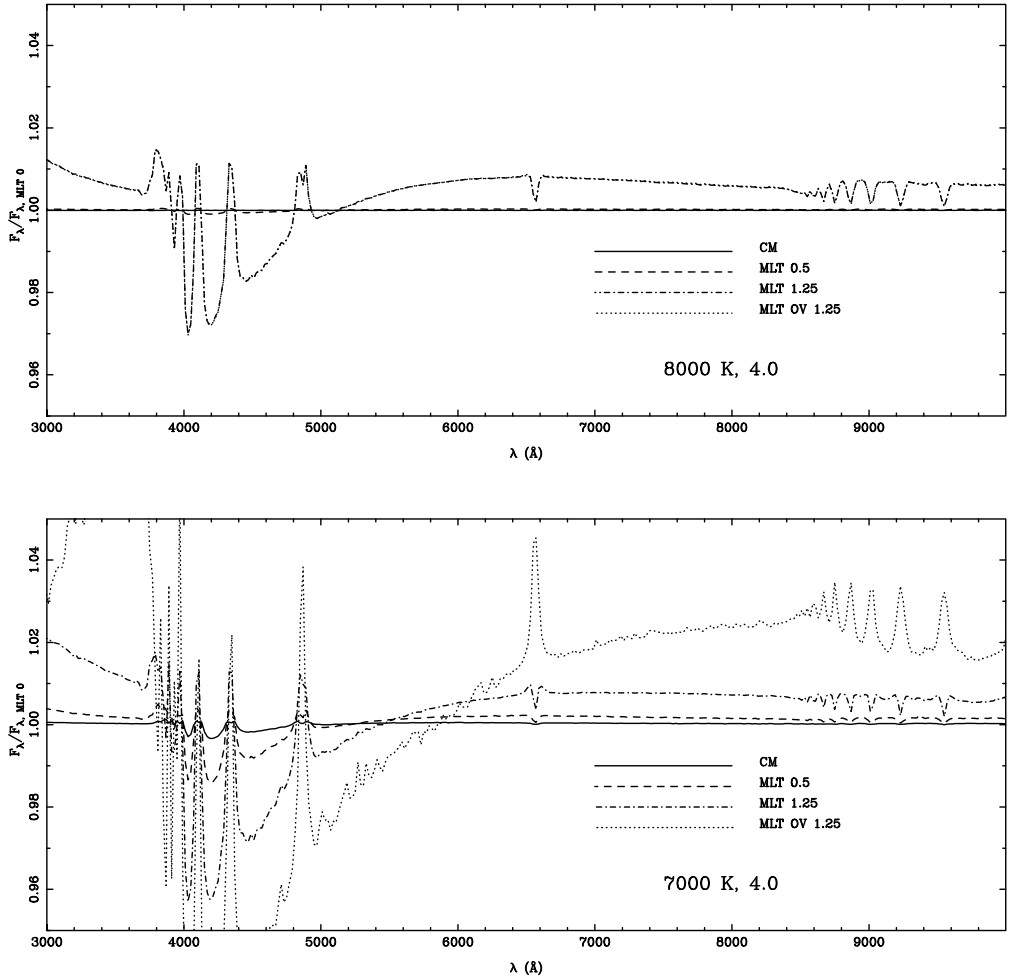}
 \caption{Fluxes for CM models and MLT models, with $l/H$ = 0.5 and 1.25,
 compared to that for a model with zero convection. At 8000~K the CM and MLT
 $l/H$ = 0.5 models give essentially the same fluxes as zero convection. Note
 that the fluxes for MLT $l/H$ = 1.25 with and without overshooting are almost
 identical. At 7000~K the differences are more noticeable, especially in the
 region 4000 $\sim$ 5000\AA, and the effect of overshooting is now
 considerable.}
 \label{fig:fluxes}
\end{figure}

Figure~\ref{fig:fluxes} shows the effects of changing mixing length from 0,
through 0.5 to 1.25 on the emergent flux for solar-composition models with
8000~K and 7000~K ($\log g$ = 4.0 and \mictrb = 2 kms). At 8000~K both the CM
and MLT $l/H$ = 0.5 models give essentially the same fluxes as zero convection.
At 7000~K, however, the differences are more noticeable, with the effect of
overshooting being considerable.

In the ultraviolet the effects are even more significant, but comparison with
observations would be complicated by the significant amount of metal line
blanketing in this region. In contrast, in the infrared the effects are
generally less compared to the optical. Hence, a combination of optical and
infrared fluxes should provide a good basis for fixing the effective
temperature of the star as well as testing the convective nature of the stellar
atmosphere. In fact, the IRFM (Blackwell \& Shallis \cite{BS77}, Blackwell \&
Lynas-Gray \cite{BL94}) is a very useful method for determining nearly
model-independent \teff\ for stars and has proved invaluable in our quest for
accurate atmospheric parameters.

Unfortunately, very little high-precision stellar spectrophotometry exists.
This situation will be rectified, once the ASTRA Spectrophotometer (Adelman
et al., \cite{ADE+03}) begins operation. This will allow spectrophotometry to
be added to our observational diagnostic toolkit.

\section{Balmer Profiles}

\begin{figure}
 \includegraphics{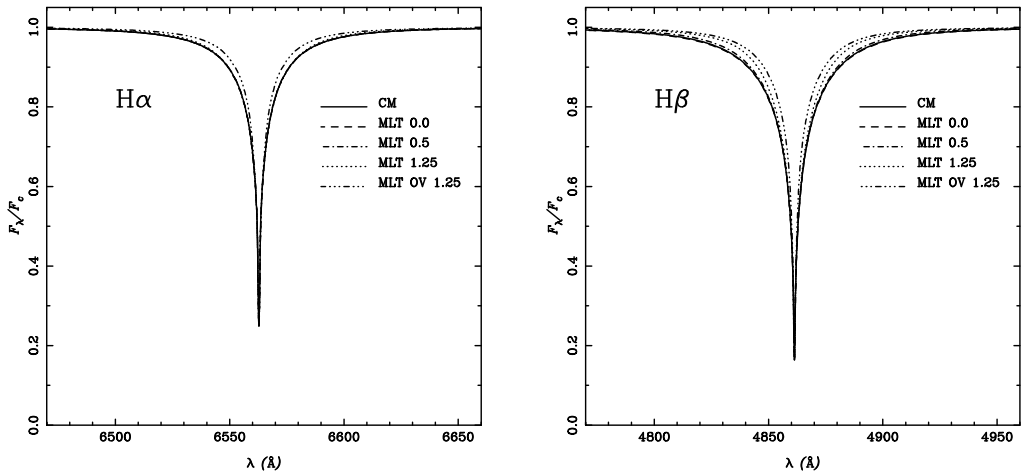}
 \caption{The effects of convection on the predicted shape of Balmer
 profiles for models with \teff\ = 7000, $\log g$ = 4.0, [M/H] = 0.0 and
 \mictrb\ = 2 \kms. $H\alpha$ is unaffected by the values of $l/H$, but
 sensitive to ``approximate overshooting'', while $H\beta$ is sensitive to
 both.}
 \label{fig:balmer}
\end{figure}

The temperature sensitivity of Balmer lines makes them an excellent diagnostic
tool for late A-type stars  and cooler (Gardiner \cite{GAR00}). However, as emphasised by van't Veer \&
M\'{e}gessier (\cite{VM96}), the $H\alpha$ and $H\beta$ profiles behave
differently due to convection: $H\alpha$ is significantly less sensitive to
mixing-length than $H\beta$. Both profiles are affected by the presence of
overshooting, with $H\beta$ being more influenced than $H\alpha$ (see
Fig.~\ref{fig:balmer}). Since $H\alpha$ is formed higher in the atmosphere than
$H\beta$, Balmer lines profiles are a very good depth probe of stellar
atmospheres. Naturally, Balmer profiles are also affected by microturbulence, metallicity and, for the
hotter stars, surface gravity (Heiter et al. \cite{HEI+02}).

In their comparison of Balmer line profiles, Gardiner et al.
(\cite{GKS99}) found that both CM and MLT without overshooting gave
satisfactory agreement with fundamental stars. Overshooting was again found to
be discrepant. Intriguingly, they found that while $l/H$ = 0.5 was generally
preferred, the region between 6000~K and 7000~K required a higher value ($l/H
\ge$ 1.25). This corresponds to the ``bump'' region found for Hyades stars
using $uvby$. However, this is not supported by either Fuhrmann et al.
(\cite{FAG93}) or van't Veer \& M\'{e}gessier (\cite{VM96}).

Gardiner et al. (\cite{GKS99}) also found evidence for significant disagreement
between all treatments of convection for stars with $T_{\rm eff}$ around 8000
$\sim$ 9000~K. Smalley et al. (\cite{SMA+02}) reviewed this region using binary
systems with known $\log g$ values and their revised fundamental $T_{\rm eff}$
values of the component stars. They found that the discrepancy found by
Gardiner et al. (\cite{GKS99}) was no longer as evident. However, this region was
relatively devoid of stars with fundamental values of both $T_{\rm eff}$ and $\log
g$. Further fundamental stars are clearly required in this region.

One potential difficulty with Balmer profiles is there great width, which can
pose problems during the extraction and rectification of observations. Locating
the `true' continuum is non-trivial and a small error can lead to significant
errors in any temperature obtained from fitting the profile (Smith \& van't Veer
\cite{SV88}).

\section{Microturbulence}

\begin{figure}
 \begin{center}
 \includegraphics{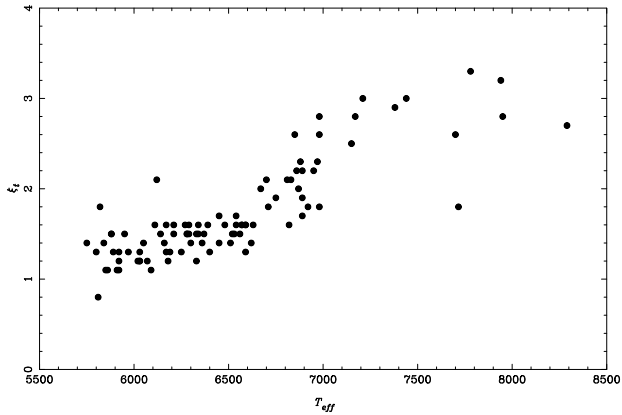}
 \end{center}
 \caption{Variation of microturbulence with effective temperature. Based on
 results of Gray (\cite{GGH01}) for stars with $\log g$ $>$ 4.0. Note the
 apparent relatively abrupt change in behaviour around 6600~K}
 \label{fig:mic}
\end{figure}

Microturbulence (long used as a free parameter in abundance analyses) does
appear to vary with effective temperature. Chaffee (\cite{CHA70}) found that
microturbulent velocity (\mictrb) rose from 2 \kms\ for early A-type stars up
to 4 \kms\ for late A-type stars, and then back to 2 \kms\ for mid F-type stars
and back up to 4 \kms\ for solar-type stars. He also found that microturbulence
correlated weakly with the Str\"{o}mgren $m_0$ index. Recently, Gray (\cite{GGH01})
found that \mictrb\ varied from around 3 \kms\ for mid-A type down to around 1
\kms\ for solar-type stars, confirming the results of Coupry \& Burkhart
(\cite{CB92}) who found a variation from around 0 \kms\ for late B-type, up to
around 3 \kms\ for mid A-type type, down to around 2 \kms\ for early F-type. Nissen
(\cite{NIS81}) gave a regression fit to both \teff\ and $\log g$ for
near-main-sequence solar-composition stars, from 2 \kms\ for F-type
down to 1.5 \kms\ for solar-type stars. Fig.~\ref{fig:mic} shows the
variation of \mictrb\ with \teff\ for near main-sequence stars
($\log g$ $>$ 4.0) based on the results given by Gray (\cite{GGH01}).

Velocity fields are present in stellar atmospheres which can be measured using
line bisectors (Dravins \cite{DRA87}, Gray \cite{GRA92}). Compared to
solar-type stars, the line bisectors are reversed, indicating small rising
columns of hot gas and larger cooler downdrafts in A-type stars (Landstreet
\cite{LAN98}).  It is these motions that are thought to be responsible, at
least in part, for the existence of microturbulence. In fact, 3d numerical
simulations of solar granulation can account for observed line profiles without
the need for any microturbulence (Asplund et al. \cite{ASP+00}). Similar
results have been found for Procyon (Gray \cite{GRA85}, Allende Prieto
\cite{AP+02}), which is also a star with well-known physical parameters (e.g.
Kervella et al. \cite{KER+04}). Unfortunately, current numerical simulations
for A-stars predict bisectors in the opposite direction to that observed
(Freytag \cite{FRE04}).

\section{B\"{o}hm-Vitense Gap}

B\"{o}hm-Vitense (\cite{BV70}) found a scarcity of stars around 7000~K which
were attributed to the abrupt onset of convection. Observationally, as
discussed above, this region corresponds to the $uvby$ ``bump'' which is a
mis-match between the $\log g$ from evolutionary models and model atmosphere
calculations, an apparent change in behaviour of microturbulence around 6600K,
and the suggestions that Balmer profiles require a higher value for
mixing-length than other regions. In addition, there is a dip in chromospheric
activity in this region (B\"{o}hm-Vitense, \cite{BV95}). D'Antona et al.
(\cite{D'A+02}) discussed the B\"{o}hm-Vitense gaps and the role of turbulent
convection, concluding that a gap in the region just below \teff\ of 7000~K is
a \teff\ rather than just a colour gap. This is an interesting area of the H-R
Diagram where the convective core shrinks to become radiative, while envelope
and atmosphere change from radiative to convective (de Bruijne et al.,
\cite{deB+01}). Chromospheric activity indicators are another useful
diagnostic, with another transition regime around 8300~K (Simon et al.
\cite{SIM+02}). All in all, this in an observationally challenging region of
the H-R Diagram, since there are both atmospheric and internal effects which we
need to differentiate between.

\section{Conclusions}

\begin{figure}
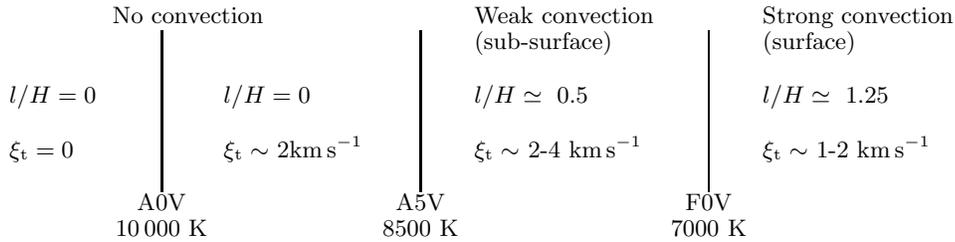


\begin{tabular*}{\textwidth}{p{1.2cm}cp{1.9cm}cp{2.4cm}cp{2.6cm}}
\multicolumn{3}{c}{No convection} && Weak convection && Strong convection \\
 &\vline&&\vline& (sub-surface) &\vline& (surface) \\
 &\vline&&\vline&&\vline& \\
$l/H = 0$ &\vline& $l/H = 0$ &\vline& $l/H \simeq\ 0.5$
&\vline& $l/H \simeq\ 1.25$ \\
 &\vline&&\vline&&\vline& \\
$\xi_{\rm t} = 0$ &\vline& $\xi_{\rm t} \sim$ 2km\,s$^{-1}$
&\vline& $\xi_{\rm t} \sim$ 2-4 km\,s$^{-1}$
&\vline& $\xi_{\rm t} \sim$ 1-2 km\,s$^{-1}$ \\
 &\vline&&\vline&&\vline& \\
&A0V&&A5V&&F0V& \\
&10\,000~K&&8500~K&&7000~K \\
\end{tabular*}
 \caption{Schematic summary of typical values of mixing-length and
 microturbulence found in the literature for various spectral types.}
 \label{fig:summary}
\end{figure}

Fig.~\ref{fig:summary} summarizes the typical results of the various
observational tests. While there are some contradictions and anomalies the
figures does illustrate the current state of our observational understanding of
stellar convection. For stars hotter than A0 there is no convection or
significant microturbulence. For the early A-type stars there is essentially no
convection within the atmosphere, since the temperature gradient is radiative,
but there are velocity fields as indicated by the modest microturbulence
values. Velocity fields increase as we go through mid to late A-type stars, and
inefficient convection is required within the atmosphere. Once convection
becomes efficient (F-type and later) the value of microturbulence is found to
drop, while the mixing-length is required to increase (although some evidence
points against this, such as the Balmer-line results discussed earlier).

Overshooting is still an issue to be resolved, since there are clearly velocity
fields above the convection zone. While the ``approximate overshooting'' of
Kurucz appears to have been discounted by observational tests, there is clearly
the need for some sort of overshooting to be incorporated within model
atmosphere calculations.

The effects of convection on the stellar atmospheric structure can be
successfully probed using a variety of observational diagnostics. The
combination of photometric colours and Balmer-line profiles has given us a
valuable insight into the nature of convection in A-type stars. High quality
observations that are currently available and those that will be in the near
future, will enable further refinements in our theoretical models of convection
and turbulence in stellar atmospheres.

\begin{acknowledgments}

The author gratefully acknowledges the support of PPARC, the Royal Society and
the IAU. This research has made use of model atmospheres calculated with {\sc
atlas9} at the Department of Astronomy of the University of Vienna, Austria,
available from {\tt http://ams.astro.univie.ac.at/nemo/}.

\end{acknowledgments}

\begin{discussion}

\discuss{Grevesse}{Turbulence in the Sun not only shows up in the asymmetries of
the lines through the shapes of the bisectors but also in a systematic but small
blue shift of the centre of the line which is related to the depth of formation
of the line.}

\discuss{Kupka}{In stellar evolution, ``alpha'' is tuned to match the radius of
the present sun. Now this may be calibrated by numerical simulations. But such
an MLT calibrated model such matches the right entropy jump. It's actual
temperature gradient is likely to be wrong (and it will actually continue to
fail in helioseismological tests), as will be the radiation field, and the
colours computed from it. Thus, just tuning the alpha with simulations won't
help - you can only get one quantity at and perhaps near the calibration point.
It does not dispatch us from finding a better model, and the alpha to match a
particular quantity may be whatever in a particular case.}

\discuss{Smalley}{Indeed, from a purely observational viewpoint we could use a
negative mixing-length if that fitted the observational data better!}

\end{discussion}
\end{document}